\documentclass
[superscriptaddress,secnumarabic,amssymb,amsmath,nobibnotes,aps,prd,showkeys,showpacs,nofootinbib,onecolumn]{revtex4}%
\usepackage{graphicx}
\usepackage{epsf}
\usepackage{bm}
\usepackage{amsmath}
\usepackage{amsfonts}
\usepackage{amssymb}%
\providecommand{\U}[1]{\protect\rule{.1in}{.1in}}

\newcommand{\be}{\begin{equation}}
\newcommand{\ee}{\end{equation}}

\newcommand{\mincir}{\raise
-3.truept\hbox{\rlap{\hbox{$\sim$}}\raise4.truept\hbox{$<$}\ }}
\newcommand{\magcir}{\raise
-3.truept\hbox{\rlap{\hbox{$\sim$}}\raise4.truept\hbox{$>$}\ }}

\begin{document}
\title{Minisuperspace Quantization of $f\left(  T,B\right)  $ Cosmology}
\author{Andronikos Paliathanasis}
\email{anpaliat@phys.uoa.gr}
\affiliation{Institute of Systems Science, Durban University of Technology, Durban 4000,
South Africa}

\begin{abstract}
We discuss the quantization in the minisuperspace for the generalized
fourth-order teleparallel cosmological theory known as $f\left(  T,B\right)
$. Specifically we focus on the case where the theory is linear on the torsion
scalar, in that consideration we are able to write the cosmological field
equations with the use of a scalar field different from the scalar tensor
theories, but with the same dynamical constraints as that of scalar tensor
theories. We use the minisuperspace description to write for the first time
the Wheeler-DeWitt equation. With the use of the theory of similarity
transformations we are able to find exact solutions for the Wheeler-DeWitt
equations as also to investigate the classical and semiclassical limit in the
de Broglie -Bohm representation of quantum mechanics.

\end{abstract}
\keywords{Minisupespace; quantization; Teleparallel gravity; Wheeler-DeWitt}
\pacs{98.80.-k, 95.35.+d, 95.36.+x}
\date{\today}
\maketitle

\section{Introduction}

\label{sec1}

Modified theories of gravity have drawn the attention of cosmologists the last
years because they provide geometric mechanisms for the explanation of the
cosmological observations \cite{Teg,Kowal,Komatsu,sp1}. The common feature of
the modified theories of gravity is the introduction of geometric invariants
in the Einstein-Hilbert action such that the new field equations admit
additional dynamical terms which drive the dynamics in order to explain the
observations. The most simple modification of the Einstein-Hilbert Action
which has been proposed in the literature is the $R^{2}$-gravity in which the
quadratic Ricci scalar term has been introduced \cite{tomita}. That specific
modification is the geometric mechanism for one of the well-known inflationary
models \cite{star,Bcot}. Generalizations of the latter modification lead to
the so-called $f\left(  R\right)  $-theory in which the gravitational Action
Integral is a function $f$ of the Ricci scalar \cite{Buda}. $f\left(
R\right)  $-theory is a higher-order theory while with the use of a Lagrangian
multiplier it can be written as a scalar-tensor theory
\cite{Sotiriou,odin1,farabook}. However, the use of the Ricciscalar to modify
General Relativity is not the unique approach which has been studied in the literature.

The formulation of the teleparallel equivalent of General Relativity (TEGR) is
based on the use of the curvature-less Weitzenb{\"{o}}ck connection instead of
using the torsion-less Levi-Civita connection. The Lagrangian of TEGR is
torsion scalar $T$. Contrary to the Ricciscalar for General Relativity
\cite{ein28,Hayashi79}. A generalization of TEGR gravity is the $f\left(
T\right)  $-theory \cite{Ferraro} which is inspired by the $f\left(  R\right)
$-theory. In contrary to $f\left(  R\right)  $-theory, $f\left(  T\right)
$-theory is a second-order theory, since torsion tensor includes only products
of first derivatives. However, while $f\left(  T\right)  $-theory is a
second-order as General Relativity there are various differences. However, the
Ricciscalar and the Torsion scalar are not the only invariants which have been
proposed in the literature, we refer the reader to
\cite{fg1,fg2,fg3,fg4,fg5,fg6,fg7,fg8,fg9,fg10,fg11,fg12} and references therein.

In this work we are interested in the modified theory of gravity known as
$f\left(  T,B\right)  $ theory, where $T$ is the torsion scalar and $B$ is the
boundary term defined as $B=T+R~$\cite{myr11}. Because~$B$ includes second
derivatives, $f\left(  T,B\right)  $ is a fourth-order theory of gravity. In
general is different from that of $f\left(  R\right)  $ theory. Specifically
the latter is recovered for $f\left(  T,B\right)  =f\left(  B-T\right)  $.
Moreover, \ in the simple case when $f_{,BB}=0$, the theory reduces to that of
$f\left(  T\right)  $ teleparallel gravity \cite{bahamonde}, thus in this
study we shall consider the case where $f_{,BB}\neq0$. Some recent analysis on
$f\left(  T,B\right)  $ gravity can be found in \cite{ftb01,ftb02,ftb02a}
where in \cite{ftb03} the modified theory is tested for the solution of the
$H_{0}$ tension.

A special case of the $f\left(  T,B\right)  $ theory which has been studied
before is that in which $f_{,TT}=0,~f_{,TB}=0\,,\ $which means that $f$ is a
linear function of $T$, that is, $f\left(  T,B\right)  =T+F\left(  B\right)  $
\cite{an1}. In the latter consideration in the case of
Friedmann--Lema\^{\i}tre--Robertson--Walker (FLRW) universe, the theory can be
written as a scalar field theory, but not a scalar tensor theory, with the
same number of dynamical constraints as the $f\left(  R\right)  $-theory. That
observation leads to field equations which can be described by a point-like
Lagrangian with the same number of constraint equations as $f\left(  R\right)
$-gravity which means that there is a minisuperspace description for the
theory. The general asymptotic behaviour as also the stability of some
important cosmological solutions such are the scaling or the de Sitter
solutions have been studied before in \cite{an1,an2}. Moreover, the
integrability properties of the field equations for this modified theory of
gravity were investigated in \cite{an3}.

We make use of the existence of the minisuperspace for the latter $f\left(
T,B\right)  $ theory in cosmological studies such that to perform a
quantization following the minisuperspace quantization which leads to the
Wheeler-DeWitt (WdW) equation \cite{wd1}. The WdW equation is actually in
general a hyperbolic functional differential equation on a spatial superspace
with infinite degrees of freedom. However, when there exists a minisuperspace
description the infinite degrees of freedom reduce to a finite number and the
WdW equation is represented as a single equation for all the points of the
spatial hypersurface. The WdW has been investigated for various modified
$f$-theories of gravity \cite{vak1,ss1,ss2,wdw2,wdw3} but not for a
teleparallel $f$-theory before. Recently, in the de Broglie-Bohm
representation of quantum mechanics it was found that in the semiclassical
limit of the WdW equation for the Szekeres universe the field equations are
modified by a quantum potential such that the Szekeres universe does not
remain silent in the early universe \cite{wdw4}. Moreover, in scalar field
theory, the same approach gives a mechanism in which terms of a pressureless
fluid are introduced in the field equations \cite{wdw5}. An interesting
discussion and critique on the WdW equation can be found in \cite{wdw6}.\ The
plan of the paper is as follows.

In Section \ref{sec2}, we present the cosmological model of our consideration,
we reproduce previous results and we show how the $f\left(  T,B\right)  $
theory can be written with the use of a Lagrange multiplier into a scalar
field theory with a minisuperspace description. In Section \ref{sec3}, we
write the WdW equation for the theory of our analysis and we apply the theory
of similarity transformations in order to constrain the unknown functional
form of the theory such that the similarity transformations which lead to the
existence of exact solutions. The complete classification for the similarity
transformations with generators point symmetries is presented. Furthermore,
the one-dimensional optimal system is derived. The latter is used to write all
the exact wavefunctions. Furthermore, in Section \ref{sec4} we investigate the
classical limit in the WKB approximate where we find the analytic solutions
for the classical gravitational field equations In addition we investigate the
quantum potentiality of the theory according to the Bohmian representation of
quantum mechanics. Finally in Section \ref{con00}, we summarize our results
and we draw our conclusions.

\section{$f\left(  T,B\right)  $ cosmology}

\label{sec2}

Consider ${\mathbf{e}_{i}(x^{\mu})~}$to be the vierbein fields, which are the
dynamical variables of teleparallel gravity. Vierbein fields form an
orthonormal basis for the tangent space at each point $P$ with coordinates,
$P\left(  x^{\mu}\right)  $, of the manifold. Hence, $g(e_{i},e_{j}%
)=\mathbf{e}_{i}\cdot\mathbf{e}_{i}=\eta_{ij}$, where $\eta_{ij}~$is the line
element of four-dimensional Minkowski spacetime. In a coordinate basis~the
vierbeins are expressed as $e_{i}=h_{i}^{\mu}\left(  x\right)  \partial_{i},$
from where it follows that the metric of the spacetime is expressed as
$g_{\mu\nu}(x)=\eta_{ij}h_{\mu}^{i}(x)h_{\nu}^{j}(x).$

The main characteristic of the teleparallel gravity is the curvatureless
Weitzenb\"{o}ck connection $\hat{\Gamma}^{\lambda}{}_{\mu\nu}=h_{a}^{\lambda
}\partial_{\mu}h_{\nu}^{a}$ from where we can define the nonnull torsion
tensor, \cite{ftt0,ftt1} $T_{\mu\nu}^{\beta}=\hat{\Gamma}_{\nu\mu}^{\beta
}-\hat{\Gamma}_{\mu\nu}^{\beta}=h_{i}^{\beta}(\partial_{\mu}h_{\nu}%
^{i}-\partial_{\nu}h_{\mu}^{i}).~$On the other hand, the Lagrangian density of
the teleparallel gravity, from which is the scalar $\ T={S_{\beta}}^{\mu\nu
}{T^{\beta}}_{\mu\nu},~$where ${S_{\beta}}^{\mu\nu}~$is defined as ${S_{\beta
}}^{\mu\nu}=\frac{1}{2}({K^{\mu\nu}}_{\beta}+\delta_{\beta}^{\mu}{T^{\theta
\nu}}_{\theta}-\delta_{\beta}^{\nu}{T^{\theta\mu}}_{\theta}).~{K^{\mu\nu}%
}_{\beta}$ is the contorsion tensor and equals the difference between the
Levi-Civita connections in the holonomic and the nonholonomic frame and it is
defined by the nonnull torsion tensor, ${T^{\mu\nu}}_{\beta}$, as ${K^{\mu\nu
}}_{\beta}=-\frac{1}{2}({T^{\mu\nu}}_{\beta}-{T^{\nu\mu}}_{\beta}-{T_{\beta}%
}^{\mu\nu}).$

$f\left(  T,B\right)  $ gravity is an extension of teleparallel theory.
$f\left(  T,B\right)  $ is a fourth-order theory where the Action integral is
a function of scalar $T$ and of the boundary term~$B=2e_{\nu}^{-1}%
\partial_{\nu}\left(  eT_{\rho}^{~\rho\nu}\right)  $, which is defined as
$B=T+R$, where $R$ is the Ricciscalar. Specifically, the Action integral is
defined as
\begin{equation}
S\equiv\frac{1}{16\pi G}\int d^{4}xe\left[  f(T,R+T)\right]  +S_{m}\equiv
\frac{1}{16\pi G}\int d^{4}xe\left[  f(T,B)\right]  +S_{m}, \label{ftb.01}%
\end{equation}
with $e=\det(e_{\mu}^{i})=\sqrt{-g}~$and $~S_{m}$ describes the additional
matter sources.

Variation with respect to the vierbein fields of (\ref{ftb.01}) provides the
field equations \cite{bahamonde}
\begin{align}
4\pi Ge\mathcal{T}_{a}^{\left(  m\right)  }{}^{\lambda}  &  =\frac{1}{2}%
eh_{a}^{\lambda}\left(  f_{,B}\right)  ^{;\mu\nu}g_{\mu\nu}-\frac{1}{2}%
eh_{a}^{\sigma}\left(  f_{,B}\right)  _{;\sigma}^{~~~;\lambda}+\frac{1}%
{4}e\left(  Bf_{,B}-\frac{1}{4}f\right)  h_{a}^{\lambda}\,+(eS_{a}{}%
^{\mu\lambda})_{,\mu}f_{,T}\nonumber\\
&  ~\ ~+e\left(  (f_{,B})_{,\mu}+(f_{,T})_{,\mu}\right)  S_{a}{}^{\mu\lambda
}~-ef_{,T}T^{\sigma}{}_{\mu a}S_{\sigma}{}^{\lambda\mu}, \label{ftb.02}%
\end{align}
where $\mathcal{T}_{a}^{\left(  m\right)  }{}^{\lambda}$ is the
energy-momentum tensor of the matter source. When $f\left(  T,B\right)  $ is
linear on~$B$, i.e. $f\left(  T,B\right)  =f\left(  T\right)  +f_{1}B$ the
latter equations take the form of $f\left(  T\right)  $ teleparallel gravity

With the use of the Einstein tensor $G_{a}^{\lambda}$ the field equations are
expressed as
\begin{align}
4\pi Ge\mathcal{T}_{a}^{\left(  m\right)  }{}^{\lambda}  &  =ef_{,T}%
G_{a}^{\lambda}+\left[  \frac{1}{4}\left(  Tf_{,T}-f\right)  eh_{a}^{\lambda
}+e(f_{,T})_{,\mu}S_{a}{}^{\mu\lambda}\right]  +\label{ftb.04}\\
&  +\left[  e(f_{,B})_{,\mu}S_{a}{}^{\mu\lambda}-\frac{1}{2}e\left(
h_{a}^{\sigma}\left(  f_{,B}\right)  _{;\sigma}^{~~~;\lambda}-h_{a}^{\lambda
}\left(  f_{,B}\right)  ^{;\mu\nu}g_{\mu\nu}\right)  +\frac{1}{4}%
eBh_{a}^{\lambda}f_{,B}\right] \nonumber
\end{align}
or%
\begin{equation}
ef_{,T}G_{a}^{\lambda}=4\pi Ge\mathcal{T}_{a}^{\left(  m\right)  }{}^{\lambda
}+4\pi Ge\mathcal{T}_{a}^{\left(  DE\right)  }{}^{\lambda}, \label{ftb.05}%
\end{equation}
that is,
\begin{equation}
eG_{a}^{\lambda}=G_{eff}\left(  e\mathcal{T}_{a}^{\left(  m\right)  }%
{}^{\lambda}+e\mathcal{T}_{a}^{\left(  DE\right)  }{}^{\lambda}\right)  ,
\label{ftb.07}%
\end{equation}
in which now $G_{eff}=\frac{4\pi G}{f_{,T}},$is an effective varying
gravitational constant.

We have defined as~$\mathcal{T}_{a}^{\left(  DE\right)  }{}^{\lambda}$ the
effective energy momentum tensor which attributes the additional dynamical
terms which follows from the modified Action Integral,
\begin{align}
4\pi Ge\mathcal{T}_{a}^{\left(  DE\right)  }{}^{\lambda}  &  =-\left[
\frac{1}{4}\left(  Tf_{,T}-f\right)  eh_{a}^{\lambda}+e(f_{,T})_{,\mu}S_{a}%
{}^{\mu\lambda}\right]  +\label{ftb.06}\\
&  -\left[  e(f_{,B})_{,\mu}S_{a}{}^{\mu\lambda}-\frac{1}{2}e\left(
h_{a}^{\sigma}\left(  f_{,B}\right)  _{;\sigma}^{~~~;\lambda}-h_{a}^{\lambda
}\left(  f_{,B}\right)  ^{;\mu\nu}g_{\mu\nu}\right)  +\frac{1}{4}%
eBh_{a}^{\lambda}f_{,B}\right]  .\nonumber
\end{align}

The geometric energy momentum tensor reads $\mathcal{T}_{a}^{\left(
DE\right)  }{}^{\lambda}=\mathcal{T}_{a}^{\left(  B\right)  }{}^{\lambda
}+\mathcal{T}_{a}^{\left(  B\right)  }{}^{\lambda}$~in which \cite{an1}%
\begin{equation}
4\pi Ge\mathcal{T}_{a}^{\left(  T\right)  }{}^{\lambda}=-\left[  \frac{1}%
{4}\left(  Tf_{,T}-f\right)  eh_{a}^{\lambda}+e(f_{,T})_{,\mu}S_{a}{}%
^{\mu\lambda}\right]  \label{ftb.10}%
\end{equation}
and $\mathcal{T}_{a}^{\left(  B\right)  }{}^{\lambda}$ is given by the
expression%
\begin{equation}
4\pi Ge\mathcal{T}_{a}^{\left(  B\right)  }{}^{\lambda}=-\left[
e(f_{,B})_{,\mu}S_{a}{}^{\mu\lambda}-\frac{1}{2}e\left(  h_{a}^{\sigma}\left(
f_{,B}\right)  _{;\sigma}^{~~~;\lambda}-h_{a}^{\lambda}\left(  f_{,B}\right)
^{;\mu\nu}g_{\mu\nu}\right)  +\frac{1}{4}eBf_{,B}h_{a}^{\lambda}\right]  .
\label{ftb.11}%
\end{equation}

\subsection{The $f\left(  T,B\right)  =T+F\left(  B\right)  $ theory}

In this work we are interested in the case where $f$ is a linear function of
$T$, that is, $f\left(  T,B\right)  =T+F\left(  B\right)  $. In that case, the
only geometric fluid components which survive are the one of $\mathcal{T}%
_{a}^{\left(  B\right)  }{}^{\lambda}$ while~$G_{eff}=4\pi G$. In addition, by
using a Lagrange multiplier the extra degrees of freedom have been attributed
to a scalar field. It is important to mention that this scalar field does not
belong to the family of scalar-tensor theories \cite{farabook}.

According to the cosmological principle in large scales the universe is
isotropic and homogeneous and described by the
Friedmann--Lema\^{\i}tre--Robertson--Walker (FLRW) line element%
\begin{equation}
ds^{2}=-N^{2}\left(  t\right)  dt^{2}+a^{2}\left(  t\right)  \left(
dx^{2}+dy^{2}+dz^{2}\right)  , \label{ftb.15}%
\end{equation}
where $a\left(  t\right)  $ is the scale factor and describes the radius of
the three-dimensional Euclidean space and $N\left(  t\right)  $ is the lapse
function. Furthermore, from the cosmological principle we select the observer
to be $u^{\mu}=\frac{1}{N}\delta_{t}^{\mu}~$such that $u^{\mu}u_{\mu}=-1$.

For the vierbein we consider the following diagonal frame~$h_{\mu}%
^{i}(t)=diag\left(  N\left(  t\right)  ,a\left(  t\right)  ,a\left(  t\right)
,a\left(  t\right)  \right)  $~from which we calculate%
\begin{equation}
T=-\frac{6}{N^{2}}\left(  \frac{\dot{a}}{a}\right)  ^{2}~,~B=-\frac{6}{N^{2}%
}\left(  \frac{\ddot{a}}{a}+\frac{2\dot{a}^{2}}{a^{2}}-\frac{\dot{a}\dot{N}%
}{aN}\right)  .\, \label{ftb.15a}%
\end{equation}

We define the new variables $\phi=\Phi_{,B}\left(  B\right)  $ and $V\left(
\phi\right)  =BF\left(  B\right)  _{,B}-F\left(  B\right)  $, hence for
$N\left(  t\right)  =1$ the gravitational field equations are \cite{an1,an2}%
\begin{equation}
3H^{2}=3H\dot{\phi}+\frac{1}{2}V\left(  \phi\right)  +\rho_{m}, \label{ftb.18}%
\end{equation}%
\begin{equation}
2\dot{H}+3H^{2}=\ddot{\phi}+\frac{1}{2}V\left(  \phi\right)  -p_{m}
\label{ftb.19}%
\end{equation}
while it follows the constraint equation
\begin{equation}
\frac{1}{6}V_{,\phi}+\dot{H}+3H^{2}=0 \label{ftb.20}%
\end{equation}
where $H=\frac{\dot{a}}{a}$ is the Hubble function.

\subsubsection{Minisuperspace description}

The gravitational field equations (\ref{ftb.18})-(\ref{ftb.20}) are derived by
the point-like singular Lagrangian function \cite{an1}%
\begin{equation}
\mathcal{L}\left(  N,a,\dot{a},\phi,\dot{\phi}\right)  =-\frac{6}{N}a\dot
{a}^{2}+\frac{6}{N}a^{2}\dot{a}\dot{\phi}-Na^{3}V\left(  \phi\right)  +L_{m},
\label{ftb.21}%
\end{equation}
which is a minisuperspace description for the theory. In particular equation
(\ref{ftb.18}) follows from the variation with respect to the variable $N$,
$\frac{\partial L}{\partial N}=0$, while the rest second-order equations
follow by the variation with respect to the scale factor and the scalar field.
Moreover, we have assumed that $L_{m}$ denotes the Lagrangian component of the
additional matter source $\rho_{m},~p_{m}$. We proceed by assuming that the
additional matter source is an ideal gas, that is$~p_{m}=w_{m}\rho_{m}$ with
equation of state parameter $\dot{\rho}_{m}+3\left(  1+w_{m}\right)  H\rho
_{m}=0$, that is $\rho_{m}=\rho_{m0}a^{-3\left(  1+w_{m}\right)  }$.

Hence, from Lagrangian (\ref{ftb.21}) we can define the momentum%
\begin{equation}
p_{a}=-\frac{12}{N}a\dot{a}+\frac{6}{N}a^{2}\dot{\phi}~,~p_{\phi}=\frac{6}%
{N}a^{2}\dot{a}, \label{ftb.22}%
\end{equation}
which can be used to write the Hamiltonian of the field equations
\begin{equation}
\mathcal{H}=N\left(  \frac{p_{a}p_{\phi}}{3a^{2}}+\frac{p_{\phi}^{2}}{3}%
+a^{3}V\left(  \phi\right)  +2\rho_{m0}a^{-3w_{m}}\right)  , \label{ftb.23}%
\end{equation}
while from (\ref{ftb.18}) it follows $\mathcal{H}=0$. Moreover, the field
equations can be written in the equivalent form%
\begin{equation}
\dot{a}=N\frac{p_{\phi}}{3a^{2}}~,~\dot{\phi}=N\frac{p_{a}}{3a^{2}}+\frac
{2}{3}Np_{\phi}~,~\dot{p}_{\phi}=-Na^{3}V_{,\phi}~,
\end{equation}%
\begin{equation}
\dot{p}_{a}=N\frac{p_{a}p_{\phi}}{6a^{3}}-3a^{2}NV\left(  \phi\right)
-6\rho_{m0}Na^{-3w_{m}-1}~.
\end{equation}

The existence of the minisuperspace description for this generalized
teleparallel model is essential in order to proceed with the quantization of
the theory. Moreover because the quantum effects refer to the very early
universe we assume that there is not any contribution of the matter source in
the field equations, that is, we assume that $\rho_{m}=0$.

\section{Wheeler-DeWitt equation}

\label{sec3}

From the point-like Lagrangian (\ref{ftb.21}) we define the minisuperspace%
\begin{equation}
ds^{2}=-12ada^{2}+12a^{2}dad\phi\label{ftb.24}%
\end{equation}
which is a two-dimensional space with Ricciscalar $R_{\left(  2\right)  }=0$,
which means that it is the two-dimensional flat space.

In general, the WdW equation is defined with the use of the conformal
invariant Laplace operator $\hat{L}_{\gamma}=\Delta_{\gamma}+\frac
{n-2}{4\left(  n-1\right)  }R_{\gamma}$ where $\gamma_{ij}$ remarks the
minisuperspace metric, $\Delta_{\gamma}$ is the Laplace operator, $R_{\gamma}$
is the Ricciscalar of $\gamma$ and $n=\dim\gamma$. For the two-dimensional
minisuperspace of the theory of our consideration it follows that \ $\hat
{L}_{\gamma}=\Delta_{\gamma}$. In this case the WdW equation is equivalent to
the classical quantization $\left[  x,p\right]  =\delta_{ij}$ and replace $p$
with the operator $p=i\frac{\partial}{\partial q}$ in the Hamiltonian equation
(\ref{ftb.23}). Therefore, we write the WdW equation%
\begin{equation}
\mathcal{W}\equiv\left(  \frac{1}{3a^{3}}\left(  a\frac{\partial^{2}}{\partial
a\partial\phi}+\frac{\partial^{2}}{\partial\phi^{2}}\right)  -a^{3}V\left(
\phi\right)  \right)  \Psi\left(  a,\phi\right)  =0, \label{ftb.25}%
\end{equation}
where $\Psi\left(  a,\phi\right)  $ is the wavefunction of the universe.

In order to solve the latter partial differential equation we investigate for
specific functions of the scalar field potential $V\left(  \phi\right)  $ in
which we can define differential operators which leave the wavefunction
invariant. Specifically, we shall investigate the existence of one-parameter
point transformations which keep the WdW equation invariant. The infinitesimal
generator of the one-parameter point transformation will be called a Lie symmetry.

\subsection{Quantum operators}

Consider the vector field $\mathbf{X=}\xi^{a}\left(  a,\phi,\Psi\right)
\partial_{a}+\xi^{\phi}\left(  a,\phi,\Psi\right)  \partial_{\phi}+\eta\left(
a,\phi,\Psi\right)  \partial_{\Psi}$ defined in the jet space $\left\{
a,\phi,\Psi\right\}  $, which is the generator of the infinitesimal
one-parameter point transformation $\ P\rightarrow P^{\prime}$ defined as
\cite{Bluman,Olver}%
\begin{equation}
\left(  a^{\prime},\phi^{\prime},\Psi^{\prime}\right)  =\left(  a,\phi
,\Psi\right)  +\varepsilon\left(  \xi^{a}\left(  a,\phi,\Psi\right)
,\xi^{\phi}\left(  a,\phi,\Psi\right)  ,\eta\left(  a,\phi,\Psi\right)
\right)  ,
\end{equation}
in which $\varepsilon$ is an infinitesimal parameter. Then equation
(\ref{ftb.25}) will remain invariant under the action of the point
transformation if and only if%
\begin{equation}
\lim_{\varepsilon\rightarrow0}\frac{\mathcal{W}\left(  a^{\prime},\phi
^{\prime},\Psi^{\prime}\right)  -\mathcal{W}\left(  a,\phi,\Psi\right)
}{\varepsilon}=0 \label{ftb.26}%
\end{equation}
or equivalently $\mathcal{L}_{X}\mathcal{W=}\mu\mathcal{W}$%
,~$\operatorname{mod}\mathcal{W}=0$, where~$\mathcal{L}_{X}$ is the Lie
derivative with respect to the vector field $\mathbf{X}$ and $\mu$ is a
function which should be determined. When the latter condition is true, field
$\mathbf{X}$ is called Lie symmetry for the differential equation
$\mathcal{W}.$

For the conformal Laplace equation it was found that the generic Lie symmetry
vector $\mathbf{X}$ is of the form \cite{AnIJ1}%

\begin{equation}
\mathbf{X}=\xi^{i}\left(  y^{k}\right)  \partial_{i}+\left[  \frac{(2-n)}%
{2}\psi\Psi+\alpha_{0}\Psi+\beta\left(  a,\psi\right)  \right]  \partial
_{\Psi}\mathbf{,} \label{ftb.27}%
\end{equation}
in which $\xi^{i}\left(  y^{k}\right)  =\left(  \xi^{a}\left(  a,\phi\right)
,\xi^{\phi}\left(  a,\phi\right)  \right)  $ is a conformal Killing vector
field of the minisuperspace $\gamma_{ij}$, with conformal factor $\psi\left(
a,\phi\right)  $, that is, $\mathcal{L}_{\xi}\gamma_{ij}=2\psi\gamma_{ij}$ and
$\psi=\frac{1}{n}\nabla_{\left(  \gamma\right)  j}\xi^{j}$. Moreover, the
conformal Killing vector $\xi^{i}\left(  y^{k}\right)  $ and the effective
potential function $V_{eff}\left(  a,\phi\right)  =a^{3}V\left(  \phi\right)
$ are constraint as $\mathcal{L}_{\xi}V_{eff}+2\psi V_{eff}=0$. Moreover,
$a_{0}$ is a constant, while $\beta\left(  a,\psi\right)  $ denotes the
infinity number of solutions of the original conformal Laplace equation. These
two vector fields indicate that the differential equation is linear. The
vector field $\beta\left(  a,\phi\right)  \partial_{\Psi}$ is a trivial
symmetry vector and has not any application on the construction of similarity
solutions. Thus we shall omit it in the following analysis.

The main application of Lie point symmetries in partial differential equations
is the determination of similarity transformations which can be used to reduce
the number of indepedent variables for the equation. Indeed, we find the point
transformations in which the conformal vector field $\xi^{i}$ is written in
normal coordinates that is, we search the transformation $y^{k}\rightarrow
y^{J}$ in which the Lie symmetry vector is written in the normal
form~\cite{Bluman,Olver}%
\begin{equation}
\mathbf{X}=\xi^{i}\left(  x^{k}\right)  \partial_{i}+\left[  \frac{(2-n)}%
{2}\psi\Psi+\alpha_{0}\Psi\right]  \partial_{\Psi}\mathbf{.} \label{ftb.28}%
\end{equation}
Now there two ways to proceed with the application of the symmetry vector. The
two approaches provide the same result, that is, they are equivalent.

The first approach is the derivation of the zero-order invariants for the
symmetry vector which follow by the solution of the Lagrange system%
\begin{equation}
\frac{dy^{b}}{0}=\frac{dy^{J}}{1}=\frac{d\Psi}{\left(  \frac{2-n}{2}%
\psi+\alpha_{0}\right)  \Psi}~,~~b\neq J, \label{ftb.29}%
\end{equation}
that is~$y^{b},~\Psi\left(  y^{b},y^{J}\right)  =\Phi\left(  y^{b}\right)
\exp\left[  \int\left(  \frac{2-n}{2}\psi+\alpha_{0}\right)  dy^{J}\right]  .$
Therefore by defining $y^{b}$ to be the new independent variables and
$\Phi\left(  y^{b}\right)  $ the dependent variable we end with a new
differential equation known as reduced equation.

On the other hand, for partial differential equations every Lie symmetry is
equivalent to the Lie-B\"{a}cklund vector field $\hat{X}=\left(  \Psi
_{J}-\left(  \frac{2-n}{2}\psi+a_{0}\right)  \Psi\right)  \partial_{\Psi}$. A
symmetry vector transforms solutions under solutions; that is, if $\Psi$ is a
solution then $\hat{X}\Psi=a_{1}\Psi$. from where there is defined the quantum
operator~%
\begin{equation}
\Psi_{J}-\left(  \frac{2-n}{2}\psi+\alpha_{0}\right)  \Psi=\alpha_{1}\Psi.
\label{ftb.30}%
\end{equation}
which provides~$\Psi\left(  y^{b},y^{J}\right)  =\Phi\left(  y^{b}\right)
\exp\left[  \int\left(  \frac{2-n}{2}\psi+\alpha\right)  dy^{J}\right]  $,
with $\alpha=\alpha_{0}+\alpha_{1}$. Hence, it is clear that the approaches
are equivalent. Moreover, for our consideration in which $n=2$, it follows
that the quantum operator (\ref{ftb.30}) reads $\Psi_{J}-\alpha\Psi=0$, which
provides the reduction $\Psi\left(  y^{b},y^{J}\right)  =\Phi\left(
y^{b}\right)  \exp\left(  \alpha y^{J}\right)  $.

We apply the symmetry condition (\ref{ftb.26}) for equation (\ref{ftb.25}) and
we find the following functional forms of $V\left(  \phi\right)  $ in which
there exist Lie symmetries which keep the wavefunction $\Psi~$invariant.

The scalar field potentials are derived to be%
\begin{align}
V_{I}\left(  \phi\right)   &  =V_{0}e^{-\lambda\phi}~,~\label{ftb.31}\\
V_{II}\left(  \phi\right)   &  =V_{0}\left(  e^{\phi}+V_{1}e^{\left(
1+\kappa\right)  \phi}\right)  ^{-2-\frac{6}{\kappa}}e^{\frac{5+\left(
\kappa+1\right)  ^{2}}{\kappa}\phi}. \label{ftb.32}%
\end{align}

The Lie symmetries for the WdW equation for the scalar field potential
$V_{I}\left(  \phi\right)  $ are%
\begin{equation}
X_{1}=a^{-\lambda}e^{\lambda\phi}\partial_{\phi}~,~X_{2}=\lambda a\partial
_{a}+\partial_{\phi}~,~X_{3}=a^{\lambda-6}\left(  a\partial_{a}+\partial
_{\phi}\right)  ~, \label{ftb.33}%
\end{equation}
while for the potential function $V_{II}\left(  \phi\right)  $ the Lie
symmetry vector is
\begin{equation}
X_{4}=V_{1}a^{1+\kappa}\partial_{a}+a^{\kappa}\left(  V_{1}+e^{-\kappa\phi
}\right)  \partial_{\phi}~, \label{ftb.34}%
\end{equation}
while in both cases the WdW equation admits the trivial symmetry vector
$X_{\Psi}=\Psi\partial_{\Psi}$.

From the symmetry vectors we can construct the corresponding operators%
\begin{align}
\hat{Q}_{1}  &  =a^{-\lambda}e^{\lambda\phi}\frac{\partial}{\partial\phi
}~,\label{ftb.35}\\
\hat{Q}_{2}  &  =\lambda a\frac{\partial}{\partial a}+\frac{\partial}%
{\partial\phi}~,\label{ftb.36}\\
\hat{Q}_{3}  &  =a^{\lambda-6}\left(  a\frac{\partial}{\partial\alpha}%
+\frac{\partial}{\partial\phi}\right)  ~,\label{ftb.37}\\
\hat{Q}_{4}  &  =V_{1}a^{1+\kappa}\frac{\partial}{\partial a}+a^{\kappa
}\left(  V_{1}+e^{-\kappa\phi}\right)  \frac{\partial}{\partial\phi}~.
\label{ftb.38}%
\end{align}

For the potential $V_{I}\left(  \phi\right)  $ the WdW equation admits more
than one quantum operators, thus the natural question which follows is, how
many independent quantum operators can be constructed. Indeed, if we consider
general linear operator $\hat{Q}=\rho_{1}\hat{Q}_{1}+\rho_{2}\hat{Q}_{2}%
+\rho_{3}\hat{Q}_{3}$ we should define the values of the coefficients
$\rho_{1},~\rho_{2},~\rho_{3}$ which lead to independent similarity solutions.
The problem is equivalent with the derivation of the one-dimensional optimal
system for the WdW equation (\ref{ftb.25}).

By definition, for the three-dimensional Lie algebra $G_{3}$ with elements
$\left\{  X_{1},~X_{2},X_{3}\right\}  ~$and structure constants $C_{BC}^{A}$,
we define the two symmetry vectors \cite{Bluman,Olver}
\begin{equation}
\mathbf{Z}=\sum_{i=1}^{3}\rho_{i}X_{i}~,~\mathbf{Y}=\sum_{i=1}^{3}\zeta
_{i}X_{i}~,~\text{\ }\rho_{i},~\zeta_{i}\text{ are coefficient constants.}
\label{sw.04}%
\end{equation}
Then we shall say that the vector fields~$\mathbf{Z}$ and $\mathbf{Y}$ are
equivalent and provide the same similarity transformation if $\mathbf{Y}%
=\sum_{j=i}^{n}Ad\left(  \exp\left(  \epsilon_{i}X_{i}\right)  \right)
\mathbf{Z~}$or~$W=cZ~,~c=const$ that is $\zeta_{i}=c\rho_{i}$. The
operator~$Ad\left(  \exp\left(  \epsilon X_{i}\right)  \right)  X_{j}$ is
defined as
\begin{equation}
Ad\left(  \exp\left(  \epsilon X_{i}\right)  \right)  X_{j}=X_{j}%
-\epsilon\left[  X_{i},X_{j}\right]  +\frac{1}{2}\epsilon^{2}\left[
X_{i},\left[  X_{i},X_{j}\right]  \right]  +... \label{sw.07}%
\end{equation}
and it is called the adjoint representation, which has the property $Ad\left(
\exp\left(  \epsilon X\right)  \right)  X=X.$ Therefore, the derivation of all
the independent Lie symmetries and their independent linear combination lead
to the one-dimensional optimal system.

For the Lie algebra $G_{3}$ we calculate the Adjoint-representations%
\begin{equation}
Ad\left(  \exp\left(  \epsilon X_{1}\right)  \right)  X_{2}=X_{2}%
-\epsilon\lambda\left(  \lambda-6\right)  X_{1}~,~Ad\left(  \exp\left(
\epsilon X_{1}\right)  \right)  X_{3}=X_{3}~, \label{ftb.41}%
\end{equation}%
\begin{equation}
Ad\left(  \exp\left(  \epsilon X_{2}\right)  \right)  X_{1}=e^{\epsilon
\lambda\left(  \lambda-6\right)  }X_{1}~,~Ad\left(  \exp\left(  \epsilon
X_{2}\right)  \right)  X_{3}=e^{-\epsilon\lambda\left(  \lambda-6\right)
}X_{3}~, \label{ftb.42}%
\end{equation}%
\begin{equation}
Ad\left(  \exp\left(  \epsilon X_{3}\right)  \right)  X_{1}=X_{1}~,~Ad\left(
\exp\left(  \epsilon X_{3}\right)  \right)  X_{2}=X_{2}+\epsilon\lambda\left(
\lambda-6\right)  X_{3}. \label{ftb.43}%
\end{equation}
Consequently, the one-dimensional system consists by the one-dimensional Lie
algebras $\left\{  X_{1}\right\}  ~,~\left\{  X_{2}\right\}  ~,~\left\{
X_{3}\right\}  ~,~\left\{  X_{1}\pm X_{3}\right\}  $, from which it follows
that the quantum operators that we should consider in order to find all the
possible independent solutions are $\left\{  \hat{Q}_{1}\right\}  ~,~\left\{
\hat{Q}_{2}\right\}  ~,~\left\{  \hat{Q}_{3}\right\}  ~,~\left\{  \hat{Q}%
_{1}+\hat{Q}_{3}\right\}  \,$.

\subsection{Potential function $V_{I}\left(  \phi\right)  $}

For the scalar field potential $V_{I}\left(  \phi\right)  $, with the use of
the operator $\hat{Q}_{1}$ we define the constraint equation $\hat{Q}_{1}%
\Psi=q_{1}\Psi$, thus from the WdW equation we find
\begin{equation}
\Psi_{1}\left(  a,\phi\right)  =\Psi_{1}^{0}\exp\left(  \frac{q_{1}}{\lambda
}a^{\lambda}e^{-\lambda\phi}+\frac{3V_{0}}{q_{1}}\frac{a^{6-\lambda}%
}{6-\lambda}\right)  . \label{ftb.44}%
\end{equation}

Similarly with the use of the operator $\hat{Q}_{2},$ that is, $\hat{Q}%
_{2}\Psi=q_{2}\Psi$ we find the similarity solution%
\begin{equation}
\Psi_{2}\left(  a,\phi\right)  =a^{-\frac{q_{2}\left(  \lambda-3\right)
}{\lambda\left(  \lambda-6\right)  }}e^{\frac{q_{2}\phi}{2\left(
\lambda-6\right)  }}\left(  \Psi_{2}^{01}J_{\frac{q_{2}}{\lambda\left(
\lambda-6\right)  }}\left(  -2\sqrt{\frac{3V_{0}}{\lambda\left(
6-\lambda\right)  }}a^{3}e^{-\frac{\lambda}{2}\phi}\right)  +\Psi_{2}%
^{02}Y_{\frac{q_{2}}{\lambda\left(  \lambda-6\right)  }}\left(  -2\sqrt
{\frac{3V_{0}}{\lambda\left(  6-\lambda\right)  }}a^{3}e^{-\frac{\lambda}%
{2}\phi}\right)  \right)  ~, \label{ftb.45}%
\end{equation}
where $J,~Y$ are the Bessel functions.

Furthermore, from the constraint equation $\hat{Q}_{3}\Psi=q_{3}\Psi\,\,$\ we
derive the wavefunction%
\begin{equation}
\Psi_{3}\left(  a,\phi\right)  =\Psi_{3}^{0}\exp\left(  \frac{3V_{0}}%
{q_{3}\lambda}a^{\lambda}e^{-\lambda\phi}+\frac{q_{3}}{6-\lambda}a^{6-\lambda
}\right)  ~. \label{ftb.46}%
\end{equation}

Moreover, from the operator $\hat{Q}_{1}+\hat{Q}_{3}$ we construct the
constraint equation $\left(  \hat{Q}_{1}+\hat{Q}_{3}\right)  \Psi=q^{+}\Psi$
which with the use of the WdW equation provides the wavefunction%
\begin{equation}
\Psi_{+}\left(  a,\phi\right)  =\exp\left(  \frac{q^{+}}{\lambda
-6}a^{6-\lambda}\right)  \left(  \Psi_{+}^{01}\exp\left(  \Delta_{+}\left(
a^{\lambda}e^{-\lambda\phi}\left(  \lambda-6\right)  -a^{6-\lambda}\right)
\right)  +\Psi_{+}^{02}\exp\left(  \Delta_{-}\left(  a^{\lambda}%
e^{-\lambda\phi}\left(  \lambda-6\right)  -a^{6-\lambda}\right)  \right)
\right)  . \label{ftb.47}%
\end{equation}
where $\Delta_{\pm}=-\frac{-q^{+}\pm\sqrt{\left(  q^{+}\right)  ^{2}-12V_{0}}%
}{2\lambda\left(  \lambda-6\right)  }$.

\subsection{Potential function $V_{II}\left(  \phi\right)  $}

We continue our analysis with the derivation of the similarity solution for
the WdW equation for the scalar field potential $V_{II}\left(  \phi\right)  $.
Thus, with the use of the unique admitted operator $\hat{Q}_{4}$ we define the
constraint equation $\hat{Q}_{4}\Psi=q_{4}\Psi$. In order to write the
solution we prefer to work in normal coordinates, hence we perform the change
of variables
\begin{equation}
a=x^{-\frac{1}{\kappa}}~,~e^{-\kappa\phi}=\frac{V_{1}x}{e^{\kappa y}-x}.
\label{ftb.48}%
\end{equation}
In the new variables the WdW equation becomes
\begin{equation}
\left(  \kappa V_{1}e^{y\left(  6+\kappa\right)  }\frac{\partial^{2}}{\partial
x\partial y}+V_{1}e^{6y}\frac{\partial^{2}}{\partial y^{2}}-\kappa e^{6y}%
V_{1}\frac{\partial}{\partial y}+3V_{0}\right)  \Psi\left(  x,y\right)  =0
\label{ftb.49}%
\end{equation}
while the constraint equation is simplified in the simplest form $\left(
\frac{\partial}{\partial x}-q_{4}\right)  \Psi\left(  x,y\right)  =0$. Hence
the similarity solution is
\begin{equation}
\Psi\left(  x,y\right)  =e^{-q_{4}x}U\left(  y\right)  \label{ftb.50}%
\end{equation}
in which $U\left(  y\right)  $ solve the differential equation%
\begin{equation}
\left(  -q_{4}\kappa V_{1}e^{y\left(  6+\kappa\right)  }\frac{\partial
}{\partial y}+V_{1}e^{6y}\frac{\partial^{2}}{\partial y^{2}}-\kappa
e^{6y}V_{1}\frac{\partial}{\partial y}+3V_{0}\right)  U\left(  y\right)  =0
\label{ftb.51}%
\end{equation}
where a special solution for $q_{4}=0$ is
\begin{equation}
U\left(  y\right)  =e^{\frac{\kappa}{2}y}\left(  U_{1}J_{\frac{\kappa}{3}%
}\left(  \sqrt{\frac{V_{0}}{3V_{1}}}e^{-3y}\right)  +U_{2}Y_{\frac{\kappa}{3}%
}\left(  \sqrt{\frac{V_{0}}{3V_{1}}}e^{-3y}\right)  \right)  . \label{ftb.52}%
\end{equation}

On the other hand for $q_{4}\neq0$ for $\kappa=-6$ the closed form solution of
$U\left(  y\right)  $ is expressed in terms of Kummer's $M\left(  \alpha
,\beta,x\right)  $ and Tricomi's $U\left(  \alpha,\beta,x\right)  $ functions,
such as,
\begin{equation}
U\left(  y\right)  =e^{-6y}\left(  U_{1}M\left(  \alpha,2,q_{4}e^{-6y}\right)
+U_{2}U\left(  \alpha,2,q_{4}e^{-6y}\right)  \right)  ~,~\alpha=-1+\frac
{V_{0}}{12q_{4}V_{1}}\text{.} \label{ftb.53}%
\end{equation}

\section{Semi-classical limit}

\label{sec4}

In the Madelung representation \cite{md1} of the complex-wave function of the
universe $\Psi\left(  a,\phi\right)  =\Omega\left(  a,\phi\right)  e^{\frac
{i}{\hbar}S\left(  a,\phi\right)  }$, \ the real part of the WdW equation
(\ref{ftb.25}) reads%
\begin{equation}
\frac{1}{3a^{3}}\left(  a\left(  \frac{\partial S}{\partial a}\right)  \left(
\frac{\partial S}{\partial\phi}\right)  +\left(  \frac{\partial S}%
{\partial\phi}\right)  ^{2}\right)  +a^{3}V\left(  \phi\right)  -\frac
{\hbar^{2}}{2\Omega}\Delta_{\gamma}\left(  \Omega\right)  =0, \label{ftb.54}%
\end{equation}
where in the limit $\hbar^{2}\rightarrow0$, the Hamilton-Jacobi equation of
the gravitational field equations is recovered. The additional term,
$V_{Q}=-\frac{\hbar^{2}}{2\Omega}\Delta_{\gamma}\left(  \Omega\right)  $ which
depends on the amplitude of the wavefunction$~\Psi\left(  a,\phi\right)  $ is
called the quantum potential in the de Broglie-Bohm representation of quantum
mechanics \cite{bm1,bm2}.

We continue our analysis by studying first the classical limit of the of the
WKB approximation, while secondly we investigate the case in which the effects
of the quantum potential are assumed nonzero.

\subsection{Classical limit}

In this section we study the classical limit without any quantum potential
term and we derive the solution of the Hamilton-Jacobi equation. The latter is
used to simplify the field equations.

\subsection{Potential function $V_{I}\left(  \phi\right)  $}

For the potential function $V_{I}\left(  \phi\right)  $ from the closed-form
solutions of the WdW equation we derived before we can easily see that the
solution of the Hamilton-Jacobi equation (\ref{ftb.54}) is
\begin{equation}
S\left(  a,\phi\right)  =\frac{1}{\lambda}a^{\lambda}e^{-\lambda\phi}%
+3V_{0}\frac{a^{6-\lambda}}{6-\lambda}.
\end{equation}
Hence, we calculate $p_{\phi}=-a^{\lambda}e^{-\lambda\phi},~$ $p_{a}%
=a^{\lambda-1}e^{-\lambda\phi}+3V_{0}a^{5-\lambda}$ where the field equations
are reduced to the following system of first-order ordinary differential
equations%
\begin{align}
\dot{a}  &  =-\frac{N}{3}a^{\lambda-2}e^{-\lambda\phi}~,\\
\dot{\phi}  &  =\frac{N}{3}\left(  a^{\lambda-3}e^{-\lambda\phi}%
+3V_{0}a^{3-\lambda}-a^{\lambda}e^{-\lambda\phi}\right)  .
\end{align}
Without loss of generality we assume $N\left(  t\right)  =-3a^{2-\lambda
}e^{\lambda\phi}$, which provides $a\left(  t\right)  =t$ and%
\begin{equation}
\dot{\phi}=\frac{t^{\lambda-3}e^{-\lambda\phi}+3V_{0}t^{3-\lambda}%
-t^{2}e^{-\lambda\phi}}{t^{\lambda-2}e^{-\lambda\phi}},
\end{equation}
which can be integrated explicitly.%

\[
\frac{d\phi}{da}=-\left(  a^{-1}+3V_{0}a^{-1}e^{\lambda\phi}-a^{2}\right)
\]

\subsection{Potential function $V_{I}\left(  \phi\right)  $}

As far as the second scalar field potential in the new coordinates $\left\{
x,y\right\}  $ is concerned the Hamilton-Jacobi equation reads%
\begin{equation}
\left(  \left(  \kappa V_{1}\left(  \frac{\partial S}{\partial x}\right)
\left(  \frac{\partial S}{\partial y}\right)  +V_{1}\left(  \frac{\partial
S}{\partial y}\right)  ^{2}\right)  -V_{0}e^{-\kappa y-6y}\right)  =0
\label{s1}%
\end{equation}
where we have replaced $N=3\left(  x^{1+\frac{3}{\kappa}}\left(  1-xe^{-\kappa
y}\right)  \right)  ^{-1}$. \ Moreover in the new coordinates we have $\dot
{x}=p_{y}~,~\dot{y}=\kappa V_{1}p_{x}+2V_{1}p_{y}.$

From (\ref{s1}) we derive the solution for the Hamilton-Jacobi equation
\[
S\left(  x,y\right)  =\frac{I_{0}}{2}\left(  2x-\kappa y\right)  -\frac
{\sqrt{\kappa^{2}V_{1}I_{0}^{2}+4V_{1}V_{0}e^{Ky}}}{V_{1}K}+\frac{1}{2}%
\ln\left(  \frac{\kappa V_{1}I_{0}+\sqrt{\kappa^{2}V_{1}I_{0}^{2}+4V_{1}%
V_{0}e^{Ky}}}{\kappa V_{1}I_{0}-\sqrt{\kappa^{2}V_{1}I_{0}^{2}+4V_{1}%
V_{0}e^{Ky}}}\right)  ,
\]
where $K=-\left(  6+\kappa\right)  $ and $I_{0}$ is the conservation law
corresponding to the symmetry vector $X_{4}$. Therefore%
\begin{align}
\dot{x}  &  =\frac{2V_{0}e^{Ky}}{\kappa V_{1}V_{0}-\sqrt{V_{1}\left(
\kappa^{2}I_{0}^{2}V_{1}\right)  +4V_{0}e^{Ky}}}~,~\\
\dot{y}  &  =\kappa V_{1}I_{0}+\frac{4V_{0}V_{1}e^{Ky}}{\kappa V_{1}%
V_{0}-\sqrt{V_{1}\left(  \kappa^{2}I_{0}^{2}V_{1}\right)  +4V_{0}e^{Ky}}}.
\end{align}

In the simple case where $I_{0}=0$ the Hamilton-Jacobi equation provides
$S\left(  x,y\right)  =-\frac{2}{KV_{1}}\sqrt{V_{1}V_{0}e^{Ky}},~$from which
the reduced system follows,
\begin{equation}
\dot{x}=-\sqrt{\frac{V_{0}}{V_{1}}}e^{\frac{K}{2}y}~,~\dot{y}=2\sqrt
{V_{1}V_{0}}e^{\frac{K}{2}y}%
\end{equation}
with closed-form solution $y\left(  t\right)  =-\frac{1}{K}\ln\left(
V_{0}V_{1}K^{2}\left(  t-t_{0}\right)  ^{2}\right)  $ and $x\left(  t\right)
=-\frac{1}{KV_{1}}\ln\left(  t-t_{0}\right)  +x_{0}$.

\subsection{Quantum potentiality}

For the derivation of the semi-classical solution in the de Broglie-Bohm
representation of quantum mechanics, in the wavefunction, $\Psi\left(
a,\phi\right)  =\Omega\left(  a,\phi\right)  e^{\frac{i}{\hbar}S\left(
a,\phi\right)  }$, $S\left(  a,\phi\right)  $ it is assumed to be the solution
of the modified Hamilton-Jacobi equation, which is used to reduce the field
equations into a system of two first-order ordinary differential equations.

\subsection{Potential function $V_{I}\left(  \phi\right)  $}

For the wavefunctions which correspond to $V_{I}\left(  \phi\right)  $ a
nonconstant amplitude $\Omega\left(  a,\phi\right)  $ follows from the
wavefunction $\Psi_{2}\left(  a,\phi\right)  $ as expressed by equation
(\ref{ftb.45}). For $\Psi_{2}^{02}=0$ and in the limit in which $a^{3}%
e^{-\frac{\lambda}{2}\phi}\rightarrow\infty$, the wavefunction is approximated
by $\Psi_{2}\left(  a,\phi\right)  =\sqrt{\frac{2}{\pi Z}}e^{\frac{q_{2}\phi
}{2\left(  \lambda-6\right)  }-\frac{q_{2}\left(  \lambda-3\right)  }%
{\lambda\left(  \lambda-6\right)  }\ln a}\cos\left(  Z-\frac{q_{2}}%
{\lambda\left(  \lambda-6\right)  }\frac{\pi}{2}-\frac{\pi}{4}\right)  $ where
$Z=-2\sqrt{\frac{3V_{0}}{\lambda\left(  6-\lambda\right)  }}$, while $q_{2}$
is assumed to be an imaginary number, i.e. $q_{2}=\acute{\imath}\left\vert
q_{2}\right\vert $.

Hence, $\Omega\left(  a,\phi\right)  =\sqrt{\frac{2}{\pi Z}}$ from where we
end with the quantum potential term $V_{Q}\left(  a,\phi\right)
=\frac{\lambda\left(  \lambda-6\right)  }{96}a^{-3}$, which means that when
$a^{3}e^{-\frac{\lambda}{2}\phi}\rightarrow\infty$, a term which corresponds
to an ideal gas with equation of state parameter $w=1$, that is, a stiff fluid
component, is introduced in the field equations as a quantum correction. Now
the quantum correction is important for small values of $a$, that is when
$e^{-\frac{\lambda}{2}\phi}\rightarrow\infty$.

\subsection{Potential function $V_{II}\left(  \phi\right)  $}

In a similar approach, for potential $V_{II}\left(  \phi\right)  $ and for
$q_{4}=0$, for the wavefunction (\ref{ftb.52}) in the limit $e^{-3y}%
\rightarrow\infty$, we can define the amplitude $\Omega\left(  x,y\right)
=\Omega_{0}e^{\frac{\kappa+3}{2}y}$. Hence the quantum correction term is
derived to be $V_{Q}\left(  x,y\right)  =\frac{V_{1}}{8}\left(  R^{2}%
-9\right)  e^{6y}$; however because that is true in the limit $e^{-3y}%
\rightarrow\infty$ easily it follows that $\left(  e^{3y}\right)
^{2}\rightarrow0$, that is, the quantum potential tern can be neglected.

\section{Conclusions}

\label{con00}

In this study, we focused on the quantization of an extended higher-order
teleparallel cosmological theory. In particular we considered the so-called
$f\left(  T,B\right)  $ gravity and its special form $f\left(  T,B\right)
=T+F\left(  B\right)  $. In the latter scenario the cosmological field
equations can be described by a point-like Lagrangian with the same number of
dynamical constraints with that of scalar tensor theory. Indeed, with the use
of a Lagrange multiplier the higher-order derivatives can be attribute in a
scalar field. However, the latter is different from that of scalar-tensor theories.

Because the point-like Lagrangian of the cosmological theory has the 2+1
degrees of freedom, they are the scale factor $a\left(  t\right)  $, the
scalar field $\phi\left(  t\right)  =F_{,B}\left(  B\left(  t\right)  \right)
$ and the lapse function $N\left(  t\right)  $. Consequently, we can define
the WdW equation by quantize the Hamilton function for the point-like
Lagrangian, that is, we performed a minisuperspace quantization of the theory.
According to our knowledge, this is the first minisuperspace quantization in
modified teleparallel theories in the literature. There are some previous
studies in the literature on the minisuperspace quantization $f\left(
T\right)  $ theory, however in these studies the authors did not considered
all the degrees of freedom and the constraint equations on their quantization approach.

In order to solve the WdW equation, we applied the theory of similarity
transformations. Specifically we investigated the functional forms for the
potential $V\left(  \phi\right)  =F_{,B}B-F$, where point symmetries exists.
The latter symmetries were used for the construction of quantum operators.
Furthermore, we were able to find the classical limit for this models, that
is, we solved the gravitational field equations. Finally, we investigated the
existence of quantum corrections for the field equations in the semi-classical
limit as it is given by the de Broglie -Bohm representation of quantum mechanics.

\end{document}